\documentclass[english]{article}
\usepackage[T1]{fontenc}
\usepackage[latin9]{inputenc}
\usepackage{color}
\usepackage{textcomp}
\usepackage{esint}

\makeatletter
\newcommand{\lyxaddress}[1]{
\par {\raggedright #1
\vspace{1.4em}
\noindent\par}
}
\newenvironment{lyxcode}
{\par\begin{list}{}{
\setlength{\rightmargin}{\leftmargin}
\setlength{\listparindent}{0pt}
\raggedright
\setlength{\itemsep}{0pt}
\setlength{\parsep}{0pt}
\normalfont\ttfamily}%
 \item[]}
{\end{list}}

\makeatother

\usepackage{babel}
\begin{document}

\title{{\large Unification of some classical and quantum ideas}}

\author{Jerzy Hanckowiak}

\author{(former lecturer and research worker of Wroclaw and Zielona Gora
Universities)}

\maketitle

\lyxaddress{Poland}

\lyxaddress{EU}

\lyxaddress{e-mail: hanckowiak@wp.pl}

\date{July, 2011}

\textbf{\large Mottos:}{\large \par}
\begin{quotation}
{\large {}``What lead me more or less directly to the special theory
of relativity was the conviction that the electromotive force acting
on a body in motion in a magnetic field was nothing else but an electric
field.'' Letter to the Michelson Commemorative Meeting of the Cleveland
Physics Society (1952), as quoted by R.S.Shankland, Am J Phys 32,
16 (1964), p35, republished in A P French, Special Relativity,}{\large \par}

{\large ISBN 0177710756 }{\large \par}

{\large \textquotedbl{}One creates from nothing. If you try to create
from something you're just changing something. So in order to create
something you first have to be able to create nothing. \textquotedbl{}
-- Werner Erhar}{\large \par}

{\large {}``Rather than being restrictions on the behavior of matter,
the laws of physics are restrictions on the behavior of physicists.''--V.J.
Stenger}{\large \par}\end{quotation}
\begin{abstract}
{\large The free Fock space with corresponding - information creation
and anihilation operators - supplies a kind of extended language in
which equations for n-point information (n-pi) of classical and quantum
physics are described. In this description the space and time are
treated in a similar manner and even different reference systems are
treated in a more democratic way. The information vacuum vectors in
both the classical and quantum case are introduced. Restrictions upon
n-pi leading to complete equations are derived.}{\large \par}

{\large The paper also draws attention to the fact that averaging
or smoothing of the original quantities (filtration) is not only consistent
with the experimental capabilities of people, but it is also an important
tool to understand the reality.}{\large \par}
\end{abstract}
{\large \tableofcontents{}}{\large \par}

\section{{\large Introduction}}

\textbf{\large Motto:}{\large \par}
\begin{quotation}
{\large {}``A theory of everything (TOE) is a putative theory of
theoretical physics that fully explains and links together all known
physical phenomena, and predicts the outcome of any experiment that
could be carried out in principle.'', (\cite{Internet (2011)}). }{\large \par}
\end{quotation}
{\large In presented papar we would like to show that classical and
quantum physics in many ways implement - not in such an ambitious
maner - although perhaps more generally, the }\textit{\large unification
philosophy}{\large{} which consists in acknowledging certain }\textit{\large relationships}{\large{}
between concepts and entities previously treated in a separate way.
By this we mean that such relationships take place not only for these
disciplines separately, but it takes place between them. Let me give
you one of a more formal definitions of unification taken from computer
science:}{\large \par}

{\large Given two input terms $s$ and $t$,}\textit{\large{} unification
}{\large is the process which attempts to find a substitution that
structurally identifies $s$ and $t$. If such a substitutin exists,
we call this substitution a}\textit{\large{} unifier of $s$ and $t$}{\large .
It is possible to exist}\textbf{\large{} infinitely many unifiers}{\large .}{\large \par}

{\large I give you a few examples of my understanding of the idea
of unification in which unifiers are given by the same equation or
space or notion and so on:}{\large \par}

{\large We know very well that a resignation from excess of information
called }\textit{\large filtration}{\large{} is often associated with
emerging in the sea of elements, atoms, agents, - constituting the
system - some new, often surprising phenomena as the emergence of
certain structures, patterns and so on. It also often corresponds
to an only possible, global characteristics of the system. It looks
that giving up the unnecessary details, we obtain additional knowledge
about the system. In this case we are dealing in fact with a }\textbf{\large unificatio}\textbf{\textit{\large n}}{\large{}
of the macroscopic and the microscopic description of the same discipline
with filters or projectors as }\textbf{\textit{\large unifiers}}{\large .
It is worth noting that the unification of micro and macro description
of the system is also carried out here using the same linear equations,
with appropriate additional conditions. }{\large \par}

{\large Agreeing to the loss of information, classical and quantum
physics can be described in a linear manner with linear equations
as unifiers. It turns out that linear equations  which can describe
classical and quantum systems can be considered in the free Fock space
(FFS). The Fock space is an algebraic system used in quantum mechanics
to describe quantum states with a variable or unknown number of particles.
It also appears that classical states with incomplete information
about initial and boundary conditions (systems with loss of information)
can and are described in linear way in such a space. This means, for
example, that the supperposition principle takes place even for classical
physics and that locally smoothed exact solutions can be represented
by means of supperpositions of nonlocally smoothed solutions. }{\large \par}

{\large In FFS, classical and quantum systems which have the same
action integral (Lagrangian or Hamiltonian) can be described by n-point
information (n-pi) satisfying also identical equations. Moreover,
the basic vector |0> of FFS can be modified in such a way, $|0>\rightarrow|0>_{info}$,
that the operators appearing in the equations for the generating vector
|V> of n-pi, were right invertible operators, see (\ref{eq:3.1}).
This modification has an impact on n-pi only if there are external
forces acting on the system. Guided by the analogy with quantum theories,
we interpret vector $|0>_{info}$ as a vector describing the vacuum.
Since the vector $|0>_{info}$ does not satisfy any known equation
of classical or quantum physics, and because by means of vector $|0>_{info}$
can be build all the available local information about the system,
we assume that it describes the local information vacuum, see Eq.\ref{eq:3.1}.
See also Werner Erhard's statement from the initial mottos. }{\large \par}

\subsection{Free Fock space. Operators that create and annihilate (local) information}

\textbf{\large Motto:}{\large \par}
\begin{quotation}
\textcolor{black}{\large {}``Information is whatever constrains our
beliefs}{\large .'' (\cite{Cat (2006)}, \cite{Heller (2011)}; page
63). }{\large \par}
\end{quotation}
{\large Free, or, full, or supper - Fock space (FFS) are synonyms
used to describes space of vectors }{\large \par}

{\large 
\begin{equation}
{\normalcolor {\normalcolor {\normalcolor {\normalcolor |V>=\sum_{n=1}\int d\tilde{x}_{(n)}V(\tilde{x}_{(n)})\hat{\eta}^{\star}(\tilde{x}_{1})...\hat{\eta}^{\star}(\tilde{x}_{n})|0>}}}+V_{0}|0>}\label{eq:1.1}
\end{equation}
in which classical and quantum physics are described. In fact, we
mean the classical physics with varies averaging and smoothing operations
also called filtrations - like moving averages. }{\large \par}

{\large We use here the following denotations: $V(\tilde{x}_{(n)})$-
components of the vector |V> - are n-point functions ((n-pfs), that
we will call n-point information (n-pi)), \cite{hanck (2010)b}. They
descibe some local properties of a considered classical or quantum
system. n-pi (n-pf) $V(\tilde{x}_{(n)})$ may be related to the averages
or expectation values of products of fields as well as matrices at
points $\tilde{x}_{(n)}$. They depend on n vectors $\tilde{x}_{i};i=1,...,n$
with space-time and other components to reduce the number of additional
indices: $\tilde{x_{i}}\in S$ a set. We will assume that all components
of the vectors $\tilde{x}_{i}$ are discrete quantities. In this way,
we consider discrete and discretized continuous physical systems.
In fact, we could assume that vectors $\tilde{x}_{i}$ are matrices
and this would allow their multiplication and methods of non commutative
geometry could be applied but now it would lead to premature complications. }{\large \par}

{\large $\hat{\eta}^{\star}(\tilde{x})$ are operators in FFS (not
in $S$) indexed by values of the vector $\tilde{x}\in S$, and |0>
- a vector with 0-pi component - $V_{0}$. In total, in FFS, the products
$\hat{\eta}^{\star}(\tilde{x}_{1})...\hat{\eta}^{\star}(\tilde{x}_{n})$
of operators acting on the vector $|0>$, $\hat{\eta}^{\star}(\tilde{x}_{1})...\hat{\eta}^{\star}(\tilde{x}_{n})|0>$,
create independent base that becomes the orthonormal base when we
assume Cuntz (co)relations}{\large \par}

{\large 
\begin{equation}
\hat{\eta}(\tilde{x})\hat{\eta}^{\star}(\tilde{y})=\delta(\tilde{x}-\tilde{y})\cdot\hat{I}\label{eq:1.2}
\end{equation}
 where operator $\hat{\eta}^{\star}(\tilde{y})$ is adjoint operator
to $\hat{\eta}(\tilde{y})$, $\hat{I}$- the unit operator and $\delta$-
the Dirac's or rather Kronecker's delta. For Fock spaces in which
other relations are used, see, e.g., \cite{Turbiner (1997)}. We must
also assume that the following equalities are satisfied:}{\large \par}

{\large 
\begin{equation}
\hat{\eta}(\tilde{x})|0>=0,\quad<0|\hat{\eta}^{\star}(\tilde{x})=0\label{eq:1.3}
\end{equation}
for all values of $\tilde{x}$. }{\large \par}

{\large By analogy to similar terms in QFT, we will call corresponding
operators as }\textit{\large (local) information creating $(\hat{\eta}^{\star}(\tilde{x}))$
and (local) information annihilating $(\hat{\eta}(\tilde{x}))$ operators
at the point}{\large{} $\tilde{x}$. Local information created at the
points $\tilde{x}_{1},...,\tilde{x}_{n}$ is described by the function
$V(\tilde{x}_{(n)})$. We can say that the operator $\hat{\eta}^{\star}(\tilde{x})$
is related in some way to a measurement at the {}``point'' $\tilde{x}$.
The vector |0> whose structure, at accepted assumtions, we do not
need to know, describes the }\textit{\large local information vacuum}{\large .
In fact it may contain some aggregated (}\textit{\large nonlocal information}{\large ). }{\large \par}

{\large In other words, n-pfs V , for n=1.2...., describe properties
of the system which are related in some way to points $\tilde{x}_{(n)}\equiv(\tilde{x}_{1},...,\tilde{x}_{n})$
like moving averages or averages of products of the unique {}``field''
$\varphi[\tilde{x}_{1};\alpha]\cdot\cdot\cdot\varphi[\tilde{x}_{n};\alpha]$
, see \cite{hanck (2008)}, \cite{hanck (2010)b}, with respect to
initial and boundary conditions represented here by the symbol $\alpha$.
For this reason, the n-pfs V are denoted in Statistical Field Theory
(SFT) as $<\varphi(\tilde{x}_{1})\ldots\varphi(\tilde{x}_{n})>$ ,or,
in Quantum Field Theory (QFT), where fields are operator-valued functions,
as $<\Psi|\hat{\varphi}(\tilde{x}_{1})\ldots\hat{\varphi}(\tilde{x}_{n})|\Psi>$
and are called expectation values of products $\hat{\varphi}(\tilde{x}_{1})\ldots\hat{\varphi}(\tilde{x}_{n})$. }{\large \par}

{\large Use yet vectors (\ref{eq:1.1}) instead of linear functional
series, (\cite{hanck (2010)a}; App.A), to generate these n-pi, more
closely resembles the canonical formulation of classical and quantum
theories. In addition, we avoid assumptions about the formalities
of the generating functional series.}{\large \par}

{\large We can say that that a main objective of work is a better
understanding of the linear equations for the n-pi V described in
a vector form in FFS as:}{\large \par}

{\large 
\begin{equation}
\hat{A}|V>=|\Phi>\label{eq:1.4}
\end{equation}
where we will assume that the source vector $|\Phi>$ creates (local)
information about the system described by the vector |V>. From Eq.\ref{eq:3.1}
results that the vector $|\Phi>$ is proportional to the vacuum vector
$|0>$. This is not a surprising result because the isolated system
can be treated as if it was contained in a vacuum. In a more detailed
way we could say that the vacuum, which is described by the global
(agregated) information related to the system, see (\ref{eq:3.1}),
is simultaneously responsible for local information about the system.
Naturally, in this way of thinking - the Mach philosophy is manifested! }{\large \par}

{\large The work, in some sense, extends or rather trims the previous
author's papers like \cite{hanck (2010)a}, \cite{hanck (2010)b},
although it may be read independently. }{\large \par}

\section{{\large Restrictions on n-point information (n-point functions) leading
to complete equations}}

{\large We would like to make the process of averaging or smoothing
(filtrations) to be independent of the choice of points. This is expressed
by the equalities: }{\large \par}

{\large 
\begin{eqnarray}
 & <\varphi^{k}(\tilde{x}_{1})\varphi(\tilde{x}_{k+1})\cdot\cdot\cdot\varphi(\tilde{x}_{n})>=\nonumber \\
 & <\varphi(\tilde{x}_{1})\cdot\cdot\cdot\varphi(\tilde{x}_{k})\varphi(\tilde{x}_{k+1})\cdot\cdot\cdot\varphi(\tilde{x}_{n})>|_{\tilde{x}_{1}=...=\tilde{x}_{k}}
\end{eqnarray}
 which mean that when in the n-point information (n-pi) we substitute
field $\varphi[\tilde{x}_{1};\alpha]$ by $\varphi^{k}[\tilde{x}_{1};\alpha]$
we should get the same result as if we replaced in the (k+n)-pi -
k first variables by $\tilde{x}_{1}$. In this way, you can get equations
for n-pi which }\textbf{\large do not depend explicitly}{\large{} on
the choice of averaging or smoothing. The same remark applies to a
theory with a more general nonlinear terms leading to the n-pi like
$<N[\tilde{x}_{1};\varphi]\varphi(\tilde{x}_{2})\cdot\cdot\cdot\varphi(\tilde{x}_{n})>$with
local functional $N$ expanded in the Volterra power series, \cite{rzew (1969)}.
The problem arises when the theory has nonanalitic nonlinearities,
although using the functional calculus, you can make some progress,
\cite{hanck (2010)b}. Another important restrictions imposed on n-pi
(correlation functions) are nonnegative conditions:}{\large \par}
\begin{lyxcode}
{\large 
\begin{equation}
<\varphi(\tilde{x})\ldots\varphi(\tilde{x})>=<\varphi^{k}(\tilde{x})>\geq0\label{eq:2.2}
\end{equation}
}{\large \par}
\end{lyxcode}
{\large where the equality represents the case $\varphi=0$. In fact,
for the correlation functions, we postulate more general conditions: }{\large \par}

{\large 
\begin{equation}
<\varphi(\tilde{x}_{1})\ldots\varphi(\tilde{x}_{n})>\geq0,\; if\;\{\varphi(\tilde{x}_{1};\alpha)\ldots\varphi(\tilde{x}_{n};\alpha)\}\geq0,\: for\: all\:\alpha\label{eq:2.2'}
\end{equation}
where $\alpha$represents initial and boundary conditions, for Eq.\ref{eq:3.4'},
under which the chosen averaging or smoothing process is taken. In
the case of points $(\tilde{x}_{1},...,\tilde{x}_{n})$ among which
you can not find a pair of points which are not \textquotedbl{}close\textquotedbl{},
like in the case (\ref{eq:2.2}), the conditions (\ref{eq:2.2'})
should almost always be satisfied. Derogate from these conditions
is evidenced by the large oscillations of the small \textquotedbl{}distances''. }{\large \par}

{\large It is important that restrictions like (5) lead to complete
equations for n-pi generated by the generating vector $|V>$. In fact
this is achieved through escape to infinity: the infinite set of equations
for an infinite amount of n-pi. Hence, the closure problem arises
considered in many papers, among others in the papers of the author:}{\large \par}

{\large \cite{hanck (2008),hanck (2010)a,hanck (2010)b}. }{\large \par}

\section{{\large Equations for n-pi of classical physics with (local) information
vacuum. Quantum Mach's principle? }}

\textbf{\large Motto:}{\large \par}
\begin{quotation}
{\large \textquotedbl{}I love talking about nothing. It is the only
thing I know anything about.\textquotedbl{} -- Oscar Wilde.}{\large \par}

{\large {}``Apart from the omniscience there is nothing else.''
--Z. Jacyna-Onyszkiewicz}{\large \par}
\end{quotation}
{\large We postulate the following equations for the n-pi $<\varphi(\tilde{x}_{1})\cdots(\tilde{x}_{n})>$
which by means of the generating vector |V> can be described in the
following way:}{\large \par}

{\large 
\begin{equation}
(\hat{L}+\lambda\hat{N}+\hat{G})|V>=\hat{P}_{0}|V>+\lambda\hat{P}_{0}\hat{N}|V>\equiv|0>_{info}\label{eq:3.1}
\end{equation}
with linear operators acting in FFS: }{\large \par}

{\large 
\begin{eqnarray}
\hat{L}= & \int\hat{\eta}^{\star}(\tilde{x})L[\tilde{x};\hat{\eta}]d\tilde{x}+|0><0|=\nonumber \\
 & \int\hat{\eta}^{\star}(\tilde{x})L(\tilde{x},\tilde{y})\hat{\eta}(\tilde{y})d\tilde{x}d\tilde{y}+\hat{P}_{0}\label{eq:3.2}
\end{eqnarray}
}{\large \par}

{\large 
\begin{equation}
\hat{N}=\int\hat{\eta}^{\star}(\tilde{z})N[\tilde{z};\hat{\eta}]d\tilde{z}+\hat{P}_{0}\hat{N}\label{eq:3.3}
\end{equation}
 and }{\large \par}

{\large 
\begin{equation}
\hat{G}=\int\hat{\eta}^{\star}(\tilde{x})G(\tilde{x})\label{eq:3.4}
\end{equation}
The above operators $\hat{L},\hat{N},\hat{G},$ are expressed by operators
$\hat{\eta}^{\star},\hat{\eta}$ which satisfy the Cuntz (co)relations
(\ref{eq:1.2}). }{\large \par}

{\large Eq.\ref{eq:3.1} results immediately from the following dicrete
version of original integro-differential equation, for the field $\varphi$: }{\large \par}

{\large 
\begin{equation}
L[\tilde{x};\varphi]+\lambda N[\tilde{x};\varphi]+G(\tilde{x})=0\label{eq:3.4'}
\end{equation}
which we apply to one (first) general solution appearing in the n-pi$<\varphi(\tilde{x}_{1})\ldots\varphi(\tilde{x}_{n})>$
and from (\ref{eq:1.2}) - (\ref{eq:1.3}). Here }\textit{\large L
}{\large and }\textit{\large N }{\large are given linear and nonlinear
functionals and }\textit{\large G }{\large a given function. Equations
(\ref{eq:3.4'}) describe the subtle (fine graining) structure of
the system under consideration. However, Eq.\ref{eq:3.1} describes
the averaged (coarse grained) or smooth characteristics. In both these
equations - space and time variables - are treated in a similar way
and this feature of description can be considered as the }\textit{\large space-time
unification}{\large . A unifier in this case is the definition of
n-pi (use of multi-time or rather multi-point information). Other
approach to the space-time variables, with the time variable $t$
distinguished, is presented in Sec.4.3. }{\large \par}

{\large We assume that}{\large \par}

{\large 
\begin{equation}
L[\tilde{x};<\varphi(\bullet)\varphi(\tilde{x}_{2})\ldots\varphi(\tilde{x}_{n})>]=<L[\tilde{x};\varphi(\bullet)]\varphi(\tilde{x}_{2})\ldots\varphi(\tilde{x}_{n})>\label{eq:3.4''}
\end{equation}
 In fact, this equality can be regarded as a restriction on the linear
operator (fuctional) $L$ and/or averaging (smoothing) process <...>.
Now, with the help of Cuntz relations (\ref{eq:1.2}) and Eq.\ref{eq:3.4'}
it is easy to see that Eq.\ref{eq:3.1}takes place.}{\large \par}

{\large A small modification of the r.h.s. of Eq.\ref{eq:3.1}, compared
with similar equations given in almost all previous works, is connected
with a demand of right invertability of the operators $\hat{L}$ and
$\hat{N}$ what force us to add terms $\hat{P}_{0}$and $\lambda\hat{P}_{0}\hat{N}$
to corresponding operators, see (\ref{eq:3.2}) and (\ref{eq:3.3}).
Without such modifications we can only look for a right inverse operation
satisfying, e.g., equation}{\large \par}

{\large 
\begin{equation}
\hat{N}\hat{N}_{R}^{-1}=\hat{I}-\hat{P}_{0}\label{eq:3.4'''}
\end{equation}
 which in not literally a right inverse operation satisfying equation:}{\large \par}

{\large 
\begin{equation}
\hat{N}\hat{N}_{R}^{-1}=\hat{I}\label{eq:3.4''''}
\end{equation}
 And just this last equality leads to the emergence in the right-hand
side of Eq.(\ref{eq:3.1}) - the vector $|0>_{ph}$. The amazing thing
is that vector $|0>_{info}\sim|0>$, called the }\textit{\large local
information vacuum vector}{\large{} have to be used only for $\hat{G}\neq0$
(operator describing external field in which the system is immersed).
In order not to confuse this vector with the vector describing the
quantum vacuum, we replaced the notation used in previous work (\cite{hanck (2010)b}):
$|0>_{ph}\Rightarrow|0>_{info}$. }{\large \par}

{\large Since this possibility was caused by the transition from the
detailed (fine-grained) to the less detailed (coarse-grained) description,
we can talk about the (local) information vacuum vector as the }\textit{\large phenomenon
of emergence}{\large . The above observations also give us some insight
into the human intellectual condition: If something is hard to imagine
(no theory, no subconscious assumptions), we treat it like a vacuum! }{\large \par}

{\large For $\hat{G}=0$, we get all perturbation formulas for n-pi,
for n=1,2,..., from projected Eq.\ref{eq:3.1}:}{\large \par}

{\large 
\begin{equation}
(\hat{I}-\hat{P}_{0})(\hat{L}+\lambda\hat{N})|V>=0\label{eq:3.1'}
\end{equation}
in which vector $|0>_{info}\sim|0>$is absent - what we treat as an
additional indication to treat this vector as a vacuum (describing
vacuum). }{\large \par}

{\large In order to deepen our knowledge about vacuum, see \cite{stenger (2006)},
and for a conceptual development of the vacuum in physics, see, e.g.,
\cite{zeiger (1998)}. }{\large \par}

{\large By introducing projectors $\hat{P}_{n}$ projecting on the
consecutive terms of the expansion (\ref{eq:1.1}), see (\ref{eq:3.9}),
we can express the projection properties of operators (\ref{eq:3.2})
- (\ref{eq:3.4}) as follows:}{\large \par}

{\large 
\begin{equation}
\hat{P}_{n}\hat{L}=\hat{L}\hat{P}_{n}\label{eq:3.5}
\end{equation}
(diagonal operator), where n=0,1,2,...,}{\large \par}

{\large 
\begin{equation}
\hat{P}_{n}\hat{N}=\sum_{n<m}\hat{P}_{n}\hat{N}\hat{P}_{m}\label{eq:3.6}
\end{equation}
(upper triangular), where n=0,1,2,..., and}{\large \par}

{\large 
\begin{equation}
\hat{P}_{n}\hat{G}=\hat{G}\hat{P}_{n-1}\label{eq:3.7}
\end{equation}
(lower triangular operator), where n=1,2,.... The operator values
function $\hat{N}$ can be a polynomial functional or a more general
Volterra functional power series depending on the vector variable
$\tilde{z}$ and the operator variables $\hat{\eta}(\tilde{y})$ indexed
by the vector variable $\tilde{y}$. The operator $\hat{N}$ }{\large \par}

{\large 
\begin{eqnarray}
 & \hat{N}=\int d\tilde{z}\hat{\eta}^{\star}(\tilde{z})N[\tilde{z};\hat{\eta}]+\hat{P}_{0}\hat{N}=\nonumber \\
 & \sum_{m}\int d\tilde{z}d\tilde{y}_{(m)}N(\tilde{z};\tilde{y}_{(m)})\hat{\eta}^{\star}(\tilde{z})\hat{\eta}(\tilde{y}_{1})\cdot\cdot\cdot\hat{\eta}(\tilde{y}_{m})+\hat{P}_{0}\hat{N}\label{eq:3.6'}
\end{eqnarray}
where m+1-pfs $N(\tilde{z};\tilde{y}_{(m)})$ describe usually nonlinear
interaction among constituens of the system. }{\large \par}

{\large We have similar relation for the operator }{\large \par}

{\large 
\begin{eqnarray}
 & \hat{L}=\int d\tilde{z}\hat{\eta}^{\star}(\tilde{z})L[\tilde{z};\hat{\eta}]+\hat{P}_{0}=\nonumber \\
 & \int\hat{\eta}^{\star}(\tilde{x})L(\tilde{x},\tilde{y})\hat{\eta}(\tilde{y})d\tilde{x}d\tilde{y}+\hat{P}_{0}\label{eq:3.5'}
\end{eqnarray}
}{\large \par}

{\large As we said, the operator $\hat{N}$ is related to a nonlinear
part of the strong (not averaged) formulation of theory (the original
differential equations, (\ref{eq:3.4'}) ). An extension of the operator
$\hat{N}$ is described by the operator $\hat{P}_{0}\hat{N}$ which
we propose to choose as follows:}{\large \par}

{\large 
\begin{equation}
\hat{P}_{0}\hat{N}=\sum_{m}\hat{P}_{0}\int d\tilde{z}d\tilde{y}_{(m)}N_{0}(\tilde{z};\tilde{y}_{(m)})\hat{\eta}(\tilde{y}_{2})\cdot\cdot\cdot\hat{\eta}(\tilde{y}_{m})\label{eq:3.5''}
\end{equation}
 with undetermined functions $N_{0}(\tilde{z};\tilde{y}_{(m)})$.
This choice of the operator $\hat{N}$ is dictated by the demand (\ref{eq:3.4''''})
to be a right invertible operatore. Further constraints on this operator
may be derived from requests to computation convergence and its simplicity. }{\large \par}

{\large The operator $\hat{G}$ describes a source term with a function
$G(\tilde{x})$ correponding to the external forces, for example.
It is symtomatic that diagonal and upper triangular operators, $\hat{L},\hat{N}$,
describe an interaction or self-interaction of the constituents of
the system and that lower triangular operator, $\hat{G}$, describes
an interaction with the external world. }{\large \par}

{\large As we will see in the Sec.4, the quantum properties of systems
are also describe by the lower traingular operators. Does this mean
that the quantum properties of systems are the result of their interaction
with the rest of the world? In other words we would have here kind
of quantum Mach's principle claiming that \textquotedbl{}Local physical
laws are determined by the large-scale structure of the universe.\textquotedbl{} }{\large \par}

{\large The simplest diagonal operator is the unit operator}{\large \par}

{\large 
\begin{equation}
\hat{I}=|0><0|+\int\hat{\eta}^{\star}(\tilde{x})\hat{\eta}(\tilde{x})d\tilde{x}\label{eq:3.8}
\end{equation}
Other diagonal operators are the projectors used in formulas (\ref{eq:3.5})
- (\ref{eq:3.7}) and constructed by means of the kind of tensor product
of bra and ket vectors (outer products):}{\large \par}

{\large 
\begin{eqnarray}
 & \hat{P}_{n}=\int\hat{\eta}^{\star}(\tilde{x}_{1})\cdots\hat{\eta}^{\star}(\tilde{x}_{n})|0><0|\hat{\eta}(\tilde{x}_{n})\cdots\hat{\eta}(\tilde{x}_{1})d\tilde{x}_{(n)}=\nonumber \\
 & \int\hat{\eta}^{\star}(\tilde{x}_{1})\cdots\hat{\eta}^{\star}(\tilde{x}_{n})\left(\hat{I}-\int\hat{\eta}^{\star}(\tilde{x})\hat{\eta}(\tilde{x})d\tilde{x}\right)\hat{\eta}(\tilde{x}_{n})\cdots\hat{\eta}(\tilde{x}_{1})d\tilde{x}_{(n)}\label{eq:3.9}
\end{eqnarray}
 for n=0,1,2,..., where $\hat{P}_{0}=|0><0|$. }{\large \par}

{\large The simplest upper trangular, local operator of the type (\ref{eq:3.6'})
is the local operator}{\large \par}

{\large 
\begin{equation}
\hat{N}_{1}=\int d\tilde{x}\hat{\eta}^{\star}(\tilde{x})\cdot\hat{\eta}^{2}(\tilde{x})+\hat{P}_{0}\int d\tilde{x}\hat{\eta}(\tilde{x})\label{eq:3.6''}
\end{equation}
through which, by the exponentiation, one can build other type of
local operators }{\large \par}

{\large 
\begin{equation}
\hat{N}_{n}=\hat{N}_{1}^{n}=\int d\tilde{x}\hat{\eta}^{\star}(\tilde{x})\cdot\hat{\eta}^{n+1}(\tilde{x})+...\label{eq:3.6'''}
\end{equation}
and further}{\large \par}

{\large 
\begin{equation}
\hat{N}\equiv\hat{N}_{loc}=\sum_{n}\lambda_{n}\hat{N}_{n}=f(\hat{N}_{1})+...\label{eq:3.6''''}
\end{equation}
}{\large \par}

{\large Projectors (\ref{eq:3.9}) form a complete set of orthogonal
projectors:}{\large \par}

{\large 
\begin{equation}
\sum_{n=0}\hat{P}_{n}=\hat{I},\quad and\;\hat{P}_{m}\hat{P}_{n}=\hat{P}_{n}\delta_{mn}\label{eq:3.10}
\end{equation}
We can say that projections $\hat{P}_{n}|V>$, for n=1,2,..., provide
n-point information about the local nature of the system but the projection
$\hat{P}_{0}|V>$provides rather global, agregated information. }{\large \par}

\section{{\large Comparison with quantum field theory (QFT) and a few loose
remarks}}

\subsection{Wightman functions and operations of averaging and smooting }

{\large In the case of a system representing the Universe, or for
isolated systems, the r.hs. of Eq.\ref{eq:3.1} can be interpreted
as a vacuum, see \cite{hanck (2010)b} and the end of Sec.1.1. From
the foregoing discussion results that the (classical) vacuum contains
the global information about the Universe. Like in Quantum Field Theory
(QFT) a non-trivial structure of the vacuum ($|0>_{ph}\neq|0>$) arises
only through the nonlinear theory. On the other hand, the vacuum in
QFT is defined as a state of minimum energy and is described by a
corresponding eigenvector $|\Psi_{0}>$ of the Hamilton operator.
So these two vectors can differ from one another because they belong
to different languages, but the physical meaning can be the same if
we realize that the lack of instruments that provide local information
about the system is equivalent to the absence of any material bodies,
which may correspond to a vacuum. We can illustrate the the above
correspondence as follows: }{\large \par}

{\large 
\begin{equation}
|0>_{info}\Longleftrightarrow\left\{ vacuum\right\} \Longleftrightarrow|\Psi_{0}>\label{eq:4.1}
\end{equation}
}{\large \par}

{\large In QFT identical equations as (\ref{eq:3.1}) take place,
for vacuum expectation values of products of the field operator $\hat{\varphi}(\tilde{x})$
(Wightman functions):}{\large \par}

\textit{\large 
\begin{equation}
<\Psi_{0}|\hat{\varphi}(\tilde{x}_{1})...\hat{\varphi}(\tilde{x}_{n})|\Psi_{0}>\label{eq:4.2}
\end{equation}
}{\large where the field operator $\hat{\varphi}(\tilde{x})$ satisfies
exactly the same equations as the field $\varphi$ :}{\large \par}

{\large 
\begin{equation}
L[\tilde{x};\hat{\varphi}]+\lambda N[\tilde{x};\hat{\varphi}]+G(\tilde{x})=0
\end{equation}
 see (\ref{eq:3.4'}). }{\large \par}

{\large In this case however, $n-pfs\equiv n-pi$ are not permutationally
symmetric because }{\large \par}

{\large 
\begin{equation}
\left[\hat{\varphi}(\tilde{x}_{1}),\hat{\varphi}(\tilde{x}_{2})\right]\neq0\label{eq:4.3}
\end{equation}
 for almost all $\tilde{x}_{1}and\:\tilde{x}_{2}$. Nevertheless,
the generating vector }{\large \par}

{\large 
\begin{eqnarray}
 & {\normalcolor {\normalcolor {\normalcolor |V>=\sum_{n=1}\int d\tilde{x}_{(n)}<\Psi_{0}|\hat{\varphi}(\tilde{x}_{1})...\hat{\varphi}(\tilde{x}_{n})|\Psi_{0}>\hat{\eta}^{\star}(\tilde{x}_{1})...\hat{\eta}^{\star}(\tilde{x}_{n})|0>}}}\nonumber \\
 & +V_{0}|0>
\end{eqnarray}
generating (local) n-pi $<\Psi_{0}|\hat{\varphi}(\tilde{x}_{1})...\hat{\varphi}(\tilde{x}_{n})|\Psi_{0}>$satisfies
identical Eq.\ref{eq:3.1} like in the case of classical, permutational
symmetric n-pi $<\varphi(\tilde{x}_{1})\ldots\varphi(\tilde{x}_{n})>$.
This could be interpreted as the real unification of classical and
quantum physics with identical form of Eq.\ref{eq:3.1} as an unifier,
if the same method of solving this equation could be applied in both
cases. It is not inconceivable that this last sentence is a }\textit{\large certain
understanding of the definition of unification}{\large{} stated in
the Introduction. }{\large \par}

{\large The fact that equations are identical in the cases of classical
and quantum theory - leads us to the question: Is it possible a classical
{}``averaging'', which would lead to non-symmetric n-pi? The answer
is - yes. but ... Let us consider the non-symmetric n-pi:}{\large \par}

{\large 
\begin{equation}
<\varphi(\tilde{x}_{1})\ldots\varphi(\tilde{x}_{n})>=\int d\alpha_{(n)}\varphi[\tilde{x}_{1};\alpha_{1}]\cdot\cdot\cdot\varphi[\tilde{x}_{n};\alpha_{n}]W[\alpha_{(n)}]\label{eq:4.5}
\end{equation}
where $W[\alpha_{(n)}]=W[\alpha_{1},...,\alpha_{n}]$ , is a non-symmetric
}\textit{\large probability density}{\large{} or a }\textit{\large smearing
functional }{\large by means of which n-pi are defined. Here $\alpha_{i}$,
i=1,...,n represent {}``n'' initial or/and boundary conditions or
symmetry parameters of the same system described in classical case
by the field $\varphi$ and in quantum case by the operator $\hat{\varphi}$.
For non-symmetric $W$ we get non-symmetric n-pi $<\varphi(\tilde{x}_{1})\ldots\varphi(\tilde{x}_{n})>$.
It can not be excluded that there is a $W$, for which there is}{\large \par}

{\large 
\begin{equation}
<\varphi(\tilde{x}_{1})\ldots\varphi(\tilde{x}_{n})>=<\Psi_{0}|\hat{\varphi}(\tilde{x}_{1})...\hat{\varphi}(\tilde{x}_{n})|\Psi_{0}>\label{eq:4.6}
\end{equation}
 but the situation becomes more complicated. Instead of one ensemble,
common for any n, we have n, and, it is possible, that these ensemble
are different for each n. Canonical case}{\large \par}

{\large 
\begin{equation}
W=\delta[\alpha_{1}-\alpha_{2}]...\delta[\alpha_{1}-\alpha_{n}]W[\alpha_{1}]\label{eq:4.7}
\end{equation}
 corresponds to symmetric n-pi which have nothing in commence with
quantum n-pi $<\Psi_{0}|\hat{\varphi}(\tilde{x}_{1})...\hat{\varphi}(\tilde{x}_{n})|\Psi_{0}>$.
In the latter case we should rather expect the formulas:}{\large \par}

{\large 
\begin{equation}
<\Psi_{0}|\hat{\varphi}(\tilde{x}_{1})...\hat{\varphi}(\tilde{x}_{n})|\Psi_{0}>=\int d\hat{\alpha}\hat{\varphi}[\tilde{x}_{1};\hat{\alpha}]\cdot\cdot\cdot\hat{\varphi}[\tilde{x}_{n};\hat{\alpha}]\hat{W}_{\Psi_{0}}[\hat{\alpha}]
\end{equation}
for n=1, 2,... in which $\hat{\varphi}[\tilde{x}_{1};\hat{\alpha}]\cdot\cdot\cdot\hat{\varphi}[\tilde{x}_{n};\hat{\alpha}]$
means product of operators. }{\large \par}

\subsection{Green's functions and nonsingular operators $\hat{L}$}

{\large Green's functions are another collection of n-pi used in QFT,
which are less mathematically correct but they are more convenient
in the computation and interpretation, \cite{rzew (1969)}, \cite{vasiliev (1976)}.
These properties are due mainly to their direct relation to the condition
of causality and unitarity and the fact that they are, in contrast
to the Wightman's functions, permutation symmetrical, \cite{rzew (1969)},
\cite{vasiliev (1976)}. The equation for the generating vector |V>
for the Green's functions $G(\tilde{x}_{(n)})$ can be similarilly
written as (\ref{eq:3.1}) }{\large \par}

{\large 
\begin{equation}
(\hat{L}+\lambda\hat{N}+\hat{C})|V>=\hat{P}_{0}\hat{L}|V>+\lambda\hat{P}_{0}\hat{N}|V>\equiv|0>_{info}\label{eq:4.2.1}
\end{equation}
 with vector $|0>_{info}$ defined as in Eq.\ref{eq:3.1}. The operators
$\hat{L},\hat{N}$ have the same projective properties as in classical
case and the operator $\hat{C}$, resulting from the canonical commutation
relations, is again the lower triangular operator with the following
projection properties:}{\large \par}

{\large 
\begin{equation}
\hat{P}_{n}\hat{C}=\hat{C}\hat{P}_{n-2}\label{eq:4.2.2}
\end{equation}
 for n=2,3,...and its projections are equal to zero for $n\in\{0,1\}$,
see \cite{hanck (1992)}, Sec.7. From mathematical point of view Eq.\ref{eq:4.2.1}
expresses the fact that $|V>$is a characteristic vector (functional)
of complex probability density functional or, more colloquially, it
is a functional Fourier transform of a certain functional (exponential
function of the action integral multiplied by the purely imaginary
number $i$: ($exp\{iS[\alpha]\})$), \cite{rzew (1969)}, \cite{vasiliev (1976)}. }{\large \par}

{\large The Green's functions $G(\tilde{x}_{(n)})$ can be obtained
from the Wightman functions (\ref{eq:4.2}) by means of permutations
of their arguments and preservation only these functions with time-orderings.
To get complete Eq.\ref{eq:4.2.1}, we have to use in addition the
cannonical commutation relations imposed on the operator field $\hat{\varphi}$
which do not change the dynamical equation (33). Because of operators
can also form a vector space it is worth noting that quantization
provides an interesting example of the dynamics with constraints without
reaction forces. Could it be an expression of extraordinary subtlety
of microcosm?}{\large \par}

{\large Let us assume that operator $\hat{L}$ is a non singular operator
(quantum theory case) and we can transfor Eq.\ref{eq:4.2.1} as
\begin{eqnarray}
 & (\hat{I}+\lambda\hat{L}^{-1}\hat{N}+\hat{L}^{-1}\hat{C})|V>=\hat{L}^{-1}|0>_{info}=|0>_{info}\label{eq:4.2.3}
\end{eqnarray}
This equation leads to nontrivial perturbative solutions due to the
operator $\hat{C}$. We see it by transformation of Eq.\ref{eq:4.2.3}
into equation:}{\large \par}

{\large 
\begin{equation}
\left[\hat{I}+\lambda\left(\hat{I}+\hat{L}^{-1}\hat{C})\right)^{-1}\hat{L}^{-1}\hat{N})\right]|V>=\left(\hat{I}+\hat{L}^{-1}\hat{C})\right)^{-1}|0>_{info}\label{eq:4.2.4}
\end{equation}
Its solution can be formally presented in the form:}{\large \par}

{\large 
\begin{equation}
|V>=\left[\hat{I}+\lambda\left(\hat{I}+\hat{L}^{-1}\hat{C})\right)^{-1}\hat{L}^{-1}\hat{N})\right]^{-1}\left(\hat{I}+\hat{L}^{-1}\hat{C})\right)^{-1}|0>_{info}\label{eq:4.2.4'}
\end{equation}
}{\large \par}

{\large Thus, assuming the legitimacy of a perturbation series or
Neumann series for the first inverse appearing in the above formula,
it is easy to see that }{\large \par}

{\large 
\begin{equation}
\hat{C}=0\label{eq:4.2.5}
\end{equation}
 leads to the trivial solution}{\large \par}

{\large 
\begin{equation}
|V>=|0>_{info}\label{eq:4.2.6}
\end{equation}
In other words, in quantum theories non trivial perturbation solutions
to Eq.\ref{eq:4.2.1} comes from the lower triangular operator $\hat{C}\neq0$.
In classical theories, when $\hat{C}\equiv0$, to get non-trivial
perturbative solutions, we have to assume that operator $\hat{L}$
is a right invertible. In this case the formulas (\ref{eq:4.2.3})
- (\ref{eq:4.2.4'}) do not occur and have to be substituted by the
formulas below. }{\large \par}

{\large In physics, a situation which corresponds to the non-singular
operators $\hat{L}$ (due to $\varepsilon$-prescription or restrictions
imposed on class of possible n-pi) is usually associated with general
conditions such as causality and unitarity conditions. Sometimes,
however, even in the QFT, the operators $\hat{L}$ are such singular
that any $\varepsilon$- prescription is not able to remove this defect.
It takes place when we have too much symmetry, such as gauge symmetry.
In such cases to solve Eq.\ref{eq:4.2.1} we proceed differently:
we assume that operator $\hat{L}$ is a right invertible operator.
It means that a one side inverse operator exists, $\hat{L}_{R}^{-1}$,
such that}{\large \par}

{\large 
\begin{equation}
\hat{L}\hat{L}_{R}^{-1}=\hat{I}\label{eq:4.2.7}
\end{equation}
 With the help of this operator, Eq.\ref{eq:4.2.1} can be converted
in an equivalent manner as follows:}{\large \par}

{\large 
\begin{equation}
\left\{ \hat{I}+\hat{L}_{R}^{-1}\left[\lambda\hat{N}+\hat{C}\right]\right\} |V>=\hat{P}_{L}|V>+\hat{L}_{R}^{-1}|0>_{info}\label{eq:4.2.8}
\end{equation}
 where projector $\hat{P}_{L}=\hat{I}-\hat{L}_{R}\hat{L}$. Further
transformations may look similar, with the difference that the operator
$\hat{L}$ is substituted by $\hat{L}_{R}^{-1}$. We get, for example, }{\large \par}

{\large 
\begin{eqnarray}
 & \left[\hat{I}+\lambda\left(\hat{I}+\hat{L}_{R}^{-1}\hat{C})\right)^{-1}\hat{L}_{R}^{-1}\hat{N})\right]|V>=\nonumber \\
 & \left(\hat{I}+\hat{L}_{R}^{-1}\hat{C})\right)^{-1}\left(\hat{P}_{L}|V>+\hat{L}_{R}^{-1}|0>_{info}\right)
\end{eqnarray}
To find the projection $\hat{P}_{L}|V>$we can use the perturbation
principle, \cite{hanck (2010)a}, which means that undetermined element
of Eq.47 is identified with the linear part of original theory: }{\large \par}

{\large 
\begin{equation}
\hat{P}_{L}|V>=\hat{P}_{L}|V>^{(0)}\label{eq:4.2.10}
\end{equation}
 As a consequence, the first order approximation to vector |V> is }{\large \par}

{\large 
\begin{equation}
|V>^{(0)}=\left(\hat{I}+\hat{L}_{R}^{-1}\hat{C})\right)^{-1}\left(\hat{P}_{L}|V>^{(0)}+\hat{L}_{R}^{-1}|0>_{info}^{(0)}\right)\label{eq:4.2.11}
\end{equation}
with the vector $\hat{P}_{L}|V>^{(0)}$ chosen according to classical
or quantum physics with $\lambda=0$. Of course, in case of classical
physics we should have $\hat{C}=\hat{G},$ or $\hat{C}=0$ with $\hat{L}_{R}^{-1}$instead
of }$\hat{L}^{-1}$.

\subsection{One-time n-pi (n-point information)}

{\large To be more specific let as assume that functions}{\large \par}

{\large 
\begin{equation}
\alpha=\left(\alpha_{i},\alpha_{o}\right)\label{eq:4.3.1}
\end{equation}
 describe initial and other conditions respectively, for Eq.\ref{eq:3.4'}.
Then, introducing notation: $\tilde{x}\equiv(t,\tilde{\vec{x}})$,
we have}{\large \par}

{\large 
\begin{equation}
\varphi(\tilde{x})\equiv\varphi[\tilde{x};\alpha]\rightarrow\varphi[\tilde{x};\alpha]|_{t=0}\equiv\varphi[0,\tilde{\vec{x}};\alpha]=\alpha_{i}(\tilde{\vec{x}})\label{eq:4.3.2}
\end{equation}
where $\tilde{\vec{x}}$ contains indexes related to different filelds
and space components without time $t$. Hence we have the last equality
in Eq.\ref{eq:4.3.2}. Of course, (\ref{eq:4.3.1}), (\ref{eq:4.3.2})
do not describe the most general case. For example, $\alpha_{o}$may
depend on the time $t$ as in case of non-stationary boudary conditions
which are changing over time. Then equations for n-pi, for ensemble
with such different boundary conditions, are not so simple. Of course,
still we can use moving averages with respect to the time! Hence,
perhaps, the popularity of this type of averages. }{\large \par}

{\large Let us introduce the generating functional (not vector) }{\large \par}

{\large 
\begin{equation}
V[\eta;t]=\int\delta\alpha_{i}\delta\alpha_{o}W[\alpha_{i},\alpha_{o}]exp\left\{ i\int d\tilde{\vec{x}}\varphi[t,\tilde{\vec{x}};\alpha]\eta(\tilde{\vec{x}})\right\} \label{eq:4.3.3}
\end{equation}
for the one-time n-pi $<\varphi(t,\tilde{\vec{x}}_{1})...\varphi(t,\tilde{\vec{x}}_{n})>$.
Here and elsewhere the symbol $\int$ represents a summation rather
than integration (discrete space). $\eta$ - without hat, means a
function, not an operator as in the case $\hat{\eta}$. But square
brackets express the functional dependence of corresponding quantities,
\cite{rzew (1969)}. }{\large \par}

{\large It satisfies the Hopf's evolution equation, \cite{hopf (1952)}:}{\large \par}

{\large 
\begin{equation}
i\frac{\partial}{\partial t}V[\eta;t]+H[\eta,i\frac{\delta}{\delta\eta}]V[\eta;t]=0\label{eq:4.3.3'}
\end{equation}
 where the operator $H$ linearly depends on the function $\eta$.
To get this equation we have to describe Eq.\ref{eq:3.4'} in a form }{\large \par}

{\large 
\begin{equation}
\frac{\partial}{\partial t}\varphi(\tilde{x})+L'[\tilde{x};\varphi]+\lambda N[\tilde{x};\varphi]+G(\tilde{x})=0\label{eq:4.3.3''}
\end{equation}
and take into account the formula (\ref{eq:4.3.3}). To get a first-order
evolution equation (\ref{eq:4.3.3''}) we must increase the number
of components of the vector $\tilde{x}$. }{\large \par}

{\large For the initial time $t=0$, and from (\ref{eq:4.3.2}) and
from (\ref{eq:4.3.3}) we get }{\large \par}

{\large 
\begin{eqnarray}
 & V[\eta;0]=\int\delta\alpha_{i}\delta\alpha_{o}W[\alpha_{i},\alpha_{o}]exp\left\{ i\int d\tilde{\vec{x}}\varphi[0,\tilde{\vec{x}};\alpha]\eta(\tilde{\vec{x}})\right\} =\nonumber \\
 & \int\delta\alpha_{i}\delta\alpha_{o}W[\alpha_{i},\alpha_{o}]exp\left\{ i\int d\tilde{\vec{x}}\alpha_{i}(\tilde{\vec{x}})\eta(\tilde{\vec{x}})\right\} \equiv\nonumber \\
 & \int\delta\alpha_{i}W[\alpha_{i}]exp\left\{ i\int d\tilde{\vec{x}}\alpha_{i}(\tilde{\vec{x}})\eta(\tilde{\vec{x}})\right\} \label{eq:4.3.4}
\end{eqnarray}
 Thus we see that the generating functional $V[\eta;0]$ has a form
of fuctional Fourier transform of the }\textit{\large marginal}{\large{}
functional (distribution) $W[\alpha_{i}]\equiv\int\delta\alpha_{o}W[\alpha_{i},\alpha_{o}]$.
In other words, the effect of other conditions, for example, the stationary
boundary conditions, on the initial generating functional and, via
the Hopf's evolution equation, on the generating functional $V[\eta;t]$,
is reduced to the calculation of this integral. In the case when other
conditions and initial conditions are independent quantities:}{\large \par}

{\large 
\begin{equation}
W[\alpha_{i},\alpha_{o}]=W[\alpha_{i}]W[\alpha_{o}]\label{eq:4.3.5}
\end{equation}
 the above integration leads to identity. }{\large \par}

{\large If the smearing functional $W[\alpha_{i}]$ is presented in
the form }{\large \par}

{\large 
\begin{equation}
W[\alpha_{i}]=exp\left\{ \int d\tilde{\vec{x}}S(\alpha_{i}(\tilde{\vec{x}}))\right\} \label{eq:4.3.6}
\end{equation}
with a }\textit{\large quasi-local}{\large{} functional $S$ (S can
depend as well on derivatives), which reflects subsystem }\textit{\large quasi-independence}{\large ,
\cite{Cat (2006)}, then it satisfies Schwinger equation, \cite{rzew (1969)}.
Its vector description has the form (\ref{eq:4.2.1}). So we get a
surprising result: }\textbf{\large The functional $V[\eta;0]$ describing
the initial conditions for the Hopf's equation satisfies the Schwinger's
type equation of quantum physics considered in the space of one dimension
less,}{\large{} but evolution of equal times n-pi is described by the
generating functional $V[\eta;t]$ of classical statistical physics
satisfying Hopf's evolution equation. If, for the functional S, the
action integral is chosen, then a gauge symmetry can be introduced. }{\large \par}

{\large Formally, a solution to the Hopf's equation (\ref{eq:4.3.3'}\}
can be presented as }{\large \par}

{\large 
\begin{equation}
V[\eta;t]=exp\left\{ itH[\eta,i\frac{\delta}{\delta\eta}]\right\} V[\eta;0]\label{eq:4.3.7}
\end{equation}
 where $H$ is the Hopf's operator. This formula is apparently different
in nature from the formula (\ref{eq:4.2.4'}) with the vacuum vector
$|0>_{info}$in the right hand side. But we must remember that the
functional (vector) $V[\eta;0]$ representing the initial state of
the system is simultaneously a function of the rest of the universe
and can be represented in the form (\ref{eq:4.2.4'}). In other words,
even in classical case the initial vector of a system representing
also the rest of the world has a quantum character. It is also possible
that the impact of the Universe on a isolated system is a beyond time,
because the functional V depends on functions defined in the D-1 space. }{\large \par}

\subsection{Stationary solutions}

{\large In this case }{\large \par}

{\large 
\begin{equation}
V[\eta;t]=V[\eta;0]\label{eq:4.4.1}
\end{equation}
for all $t>0$. In result the evolutionary equation like (\ref{eq:4.3.7})
loses its meaning. The problem reduces to calculating the functional
integral (\ref{eq:4.3.4}). However, the nature itself saves the value
of the evolutionary equations. It turns out that systems pushed out
of equilibrium tend to equilibrium by themselves - like in the case
of shaken vessel with liquid. This means that we can use Eq.\ref{eq:4.3.7}
with simpler initial conditions described, for example, by Gaussian
functionals, \cite{rzew (1969),vasiliev (1976)}, and get a stationary
state in the process of solving evolutionary equations with $t\rightarrow\infty$.
Sometimes people say that computers like evolutionary equations, \cite{Na (1979)}.}{\large \par}

{\large Moreover, if the dimension of the vectors $\tilde{x}$ is
increased by one, by introducing so called the fictitiouse time $s$,
then by means of Eq.\ref{eq:4.3.7} one can also describe Quantum
Field Theory, \cite{hanck (1992)}, which further reduces the differences
between classical and quantum theories. This allows you to look at
quantum system as a classical system with the lack of detailed data
responsible for the uniqueness of the solutions. Of course, we cannot
forget that mentioned above of lack of detailed data is obtained by
integration with a complex measure.}{\large \par}

{\large So things are in spacetime with one extra time, which perhaps
is wrongly called the fictitious time. In fact, we should call it
the hidden variable or rather the }\textit{\large hidden time}{\large !.
For comparison, see \cite{Khren 2006}. To reveal to us our world
- starting with a big bang until the present day, with physical laws
allowing us to predict the future and the past, we have to go with
the hidden time up to infinity. }{\large \par}

{\large Addition to the hidden time there is the }\textit{\large hidden
probability}{\large{} which is a complex valued function giving the
usual probability by calculating the square of its module. }{\large \par}

{\large One can also use Eq.\ref{eq:3.1} with the whole benefit of
the possibilities supplied by the free Fock space. }{\large \par}

\section{{\large Final remarks}}
\begin{quotation}
{\large {}``In physics, a unified field theory (occasionally referred
to as a \textquotedbl{}uniform\textquotedbl{} field theory{[}1{]})
is a type of field theory that allows all that is usually thought
of as fundamental forces and elementary particles to be written in
terms of a single field. There is no accepted unified field theory.
It remains an open line of research. The term was coined by Einstein,
who attempted to unify the general theory of relativity with electromagnetism,
hoping to recover an approximation for quantum theory. A \textquotedbl{}theory
of everything\textquotedbl{} is closely related to unified field theory,
but differs by not requiring the basis of nature to be fields, and
also attempts to explain all physical constants of nature.''; }{\large \par}

{\large {[}http://en.wikipedia.org/wiki/Unified\_field\_theory{]}.}{\large \par}
\end{quotation}
{\large From the above equations and formulas - vector $|0>_{info}$,
which should not contain any information about considered systems,\cite{hanck (2010)a},
- in fact it contains such information (global) and this is true in
classical as well as in quantum case. It is possible that this is
a manifestation of a physical inability to create a situation which
we might call absolute nothingness. We must note, however, that by
nothingness we have here in mind rather }\textit{\large information
vacuum}{\large{} about the system which we do not identify with the
absence of that system. It is not excluded that if the system is the
whole Universe then information vacuum about the Universe - due to
lack of any measuring instruments outside of the system - can be identified
with the physical vacuum describing nothingness! }{\large \par}

{\large Presented and previous studies had added to the }\textit{\large eigenvalues
and eigenvectors philosophy of physics}{\large{} a new approach in
which creation, annihilation operators and the {}``vacuum'' vectors
still appear but are rather used for creation and annihilation of
information contained in fields. }{\large \par}

{\large As an interesting issue would be an explanation of why only
the presence of the external field represented by the operator $\hat{G}$
or the presence of the canonical commutation relations represented
by the operator $\hat{C}$ lead to the impact of information vacuum,
$|0>_{info}$, on the n-pi. Would it be a real manifestation of the
true nature of the vacuum?}{\large \par}

{\large The multitimes formalism considered mainly here and other
author papers, and of course by other peoples, \cite{Krai (1962)},
\cite{Mon (1967)}, allows better aquaint ourselves with real role
of the time in description of the Nature and choose right description
in particular cases. In my opinion the multitime formalism not only
casts a new perspective on equations for the n-pi, but also shows
opportunity to formulate a theory in the form in which temporal and
spatial variables plays a similar role. In this sense we can speak
about of a }\textit{\large unification of space and time}{\large ,
even in the Newtonian theories. In many cases, as additional advantage
of the above unification, is a possibility to derive the same equations
for n-pfs $V(\tilde{x}_{(n)})$ in the two types of averages: like
in the case of (\ref{eq:4.5}) - (\ref{eq:4.7}) and in a more practical
case of averaging with respect to variables $\tilde{x}$, see \cite{hanck (2008)},
\cite{hanck (2007)}, among which are averages to which ergodic hypothesis
can be postulated, {[}16{]}. }{\large \par}

{\large Similar remark can be made for the initial conditions for
the one-time Hopf's equation, which satisfy similar equations as the
vectors |V> generating n-pi in the D-1 dimension space. Was it was
the unification of dynamics and additional conditions? See \cite{nielsen (2008)},
for a similar idea. This picture is even deeper if as smearing functional
the action integral is used in exponential function, \ref{eq:4.3.5},
which has the same symmetry as the dynamical equations in D-1 dimension.
It is worth noting that in the description of the one-time evolution,
where time is treated differently than spatial variables, in the initial
conditions - the local information vacuum appears. As an additional
evidence that an unification of dynamics and initial conditions takes
place is that the algorithmic information content of the final state
is not much greater than that of the initial state, \cite{Teg (1996)}. }{\large \par}

{\large Due to many similarities in the description of classical and
quantum physics in the full Fock space, we can consider this space
as their unifier or - to put it simply - as an arena of their unification. }{\large \par}

{\large We would like to mention the unification called the spacetimematter
unification, see W.M. Stuckey's pappers about relational blockworld,
and the unification called p-q duality, see \cite{Far (1999)}. }{\large \par}

{\large The covariant formulation of a theory can be treated as some
kind of unification of different reference frames particularly recomended
for description of large scale systems. We add that covariant formulation
means symmetrical formulation, where by this we understand only formal
distinction, \cite{Wilczek (2008)}. From that point of view it is
interesting that quantum physics can be obtained from classical physics
by reducing the symmetry (n-pi are not permutation symmetrical, see
Sec.4.1). In other words, contrary to popular belief, classical world
is more symmetrical than the quantum!}{\large \par}

{\large The loss of information can be used perhaps to understand
the role of covariant formulation of theory used by Einstein requiring
that there be no preferred reference frames and further: \textquoteleft{}The
laws of physics must be of such a nature that they apply to systems
of reference in any kind of motion'' (1916); \cite{Brad (2005)},\cite{Stenger (2007)}.
We can hope that in the case of a more symmetrical theories the moving
averages are more smooth and can be used in the case of less precise
measurements possible only in large scale systems.}{\large \par}

\textbf{\large Appendix1:}{\large \par}

{\large {}``In modern theoretical physics particle interactions are
described by gauge theories. These theories are constructed by demanding
that symmetries in the laws of physics should be local, rather than
global, in character. {}``--Anthony Lasenby, Chris Doran, and Stephen
Gull}{\large \par}

\textbf{\large Appendix2:}{\large \par}

{\large {}``The }\textit{\large free algebra on n indeterminates}{\large{}
$X_{1},...,X_{n}$ (the construction works also for any countable
set S of \textquotedblleft{}indeterminates\textquotedblright{}), is
the algebra spanned by all linear combinations }{\large \par}

{\large 
\begin{equation}
\sum\Pi^{i_{1}...i_{n}}X_{i_{1}}...X_{i_{n}}\label{eq:5.1}
\end{equation}
of formal products of the generators $X_{i}$ , with coefficients
$\Pi^{i_{1}...i_{n}}\in K$. This algebra is denoted by $K<X_{i}>$and
is said to be freely generated by the X\textquoteright{}s. {}``--X.Bekaert
(Internet). We gave this definition to note a similar structure in
FFS formed by vectors (\ref{eq:1.1}). In this case, the indeterminates
are represented by operators $\hat{\eta}^{\star}(\tilde{x})$ with
indexes $\tilde{x}$ instead of subindex $i$. }{\large \par}

\textbf{\large Appendix3:}{\large \par}

{\large General messages are defined as}{\large \par}

\begin{equation}
\mathit{A}^{+}\doteq\sum_{n=0}^{\infty}\mathcal{A^{\mathtt{n}}}
\end{equation}
 {\large where $\mathcal{A}^{n}$are n-component words: $x_{1}\cdots x_{n}$
with letters $x_{i}\in\mathit{\mathcal{A}}$, where the set $\mathcal{A}$
is called the alphabet. It follows that the book of nature can be
described both with finite and infinite alphabets. In physics, as
letters can be used distinguished particles or rather states of particles.
In case of Fock space constituated by means of vectors (\ref{eq:1.1})
the letters are operators $\hat{\eta}*$ at particular points $\tilde{x}$,
see also \cite{Bost (2006)}. }{\large \par}

{\large The following formula is also used:}{\large \par}

{\large 
\begin{equation}
\mathcal{H^{\oplus}}=\bigoplus_{n=0}^{\infty}\mathcal{H^{\otimes}}^{n}=\mathcal{H}^{\otimes0}\oplus\mathcal{H}^{\otimes}\oplus\mathcal{H}^{\otimes2}\oplus\cdots
\end{equation}
 with symbols $\otimes,\oplus$denoting the tensor product and direct
sum. $\mathcal{H}^{\oplus}$is a general message space, with the Hilbert,
alphabet space $\mathcal{H}$. \cite{Bost (2006)}. }{\large \par}

\textbf{\large Appendix4:}{\large \par}

{\large {}``If something is systematically absent, as if as a rule
it is a \textquotedbl{}non-given\textquotedbl{}, then in order to
express this state in natural language, willy-nilly, we say that it
does not exist.'' (from Tadeusz Barto\'{s}, }\textit{\large Koniec
Prawdy Absolutnej}{\large{} \textquotedbl{}The end of absolute truth\textquotedbl{},
page 169, my translation). Can we say this also about a vacuum?? }{\large \par}

\end{document}